# Observation of the superconducting-insulating transition in a BiSr(La)CuO thin film tuned by varying the oxygen content

I Matei, Z Z Li and H Raffy[1]

Laboratoire de Physique des Solides, UMR 8502-CNRS, Université Paris-Sud, 91405 Orsay, France

E-mail: raffy @lps.u-psud.fr

**Abstract.** We have conducted a highly precise doping controlled experiment on a thin film of the single layer cuprate $Bi_2Sr_{2-x}La_xCuO_y$ (x = 0.6), in order to observe the superconducting insulating transition (SIT). $Sr^{2+}/La^{3+}$ substitution (x = 0.6) allowed us to start from a doping state in the vicinity of the SIT, which was further finely-tuned by varying the oxygen content in very small amounts through a vacuum anneal process. Measurements of the Hall effect at 300K and of the resistance vs. T, R(T), have been performed after each treatment. Approaching the SIT from the metallic side, and with decreasing T, R(T) exhibits a high T metallic behaviour down to a temperature $T_{min}$, followed by a non metallic increase (dR/dT< 0) up to a maximum at $T_p$ before the superconducting transition at $T_c$. This re-entrant behaviour in R(T), commonly observed in cuprates close to the SIT, is attributed to a phase segregation phenomenon in the presence of strong electronic correlations and inhomogeneities. We present our results extracted from R(T) and Hall effect studies in the region close to the SIT and in the insulating region. A phase diagram shows the characteristic temperatures versus doping.

## 1. Introduction

Electronic properties and superconductivity in cuprate compounds are essentially governed by their carrier concentration *p*. By decreasing the hole content from optimal doping, where the critical temperature $T_c$ is maximum, superconductivity disappears at a critical given hole content $p_c$ and for $p<p_c$ the system becomes insulating at low temperature. Interestingly *p* can be monitored in a same sample by changing the oxygen content. As shown in our previous studies [1, 2] and in ref. [3], it is possible to precisely study the influence of oxygen doping, in a reversible manner, in a single film. The temperature dependence of the resistivity is one of the normal–state property which points out to unusual electronic structure in cuprates. It was observed in $Bi_2Sr_2CaCu_2O_y$ [2, 3] and $Bi_2Sr_{1.6}La_{0.4}CuO_y$ [1] thin films, that in the vicinity of $p_c$ there is an interval of doping where the resistance curve *R(T)* exhibits successively three regions when decreasing T from 300K: a metallic decrease of *R(T)* down to a minimum at $T_{min}$, a non metallic increase of *R(T)* up to a maximum at $T_p$, and a superconducting transition at $T_c$. Furthermore the SIT does not take place at the critical value of the sheet resistance, 6.5kΩ/□, and superconductivity can still be observed above this value.

We have conducted a highly precise doping controlled experiment on a thin film of the single layer cuprate $Bi_2Sr_{2-x}La_xCuO_y$, with the motivation to revisit this reentrant resistive region above $p_c$ in order to see how the SIT occurs and how superconductivity disappears. The one layer Bi (La)–based cuprate considered in this work, under-doped by $Sr^{2+}/La^{3+}$ substitution, has a low $T_{cmax}$, 8 to 10 times smaller than $T_{cmax}$ of the two-layer one. It is known to be more disordered with a structural disorder [4] and to possess a large fluctuative region [5]. After describing our experimental techniques, we will present and discuss our results on Hall effect measurements and R(T) studies in the region of the SIT and in the insulating region.

## 2. Experimental details

The reported study was conducted on a $Bi_2Sr_{2-x}La_xCuO_{6+y}$ thin film with $x=0.6$ and with thickness $t=2700$ Å. It was prepared by single target reactive rf magnetron sputtering on a heated (720-750°C) single crystal $SrTiO_3$(100) substrate. The deposition conditions were similar to those previously described for pure *Bi-2201* thin films [6]. The composition of the films was measured by Rutherford Backscattering Spectroscopy (RBS) on a film deposited on MgO substrate. The measured value of $x$ is quite close to the nominal composition of the target. The film was characterized by X-ray diffraction studies, which allows one to show that it is single-phase, *c*-axis oriented (FWHM=0.2 degrees), epitaxially grown.

The deposit was patterned mechanically, in order to avoid resist contamination, into a strip of length $l=602$ microns and width $w = 326$ microns between voltage probes and equipped with six sputtered gold contacts for four probe transport measurements. The temperature dependence of the resistance $R(T)$ of the sample was measured inside a helium dewar using a home made dipping stick. Hall measurements were performed in an electromagnet (H=1tesla) by reversing the field.

The $Sr^{2+}/La^{3+}$ cationic substitution allow us to get a sample in an underdoped state, even after complete oxidation [4]. So, after deposition, the sample, which was ex-situ oxygenated in a pure oxygen flow at 420°C, had a critical temperature $T_c(R=0) = 6.5K$ and exhibited a $R(T)$ curve characteristic of an underdoped sample. Then to approach further the SIT, the oxygen was removed in very small quantities by low temperature vacuum anneal treatments, typically at 230°C (1 minute to 30 minutes) followed by a rapid cool down at room temperature. After each treatment, the Hall constant at 300K and $R(T)$ were measured. More than 44 steps were performed successively and reversibly across the phase diagram. The SIT region itself was explored in very small steps numbered 20 to 48.

## 3. Results and discussions

The variation versus temperature of the sheet resistance $R_s(T)$ per *Bi(La)SrCuO* layer in units of $h/4e^2$ is presented in Figure 1. The sheet resistance has been derived from the expression: $Rwt/ls$ where $s$ is the spacing between $CuO_2$ layers, given by the parameter c: $s = c/2 = 12.12$ Å extracted from X-ray diffraction spectrum.

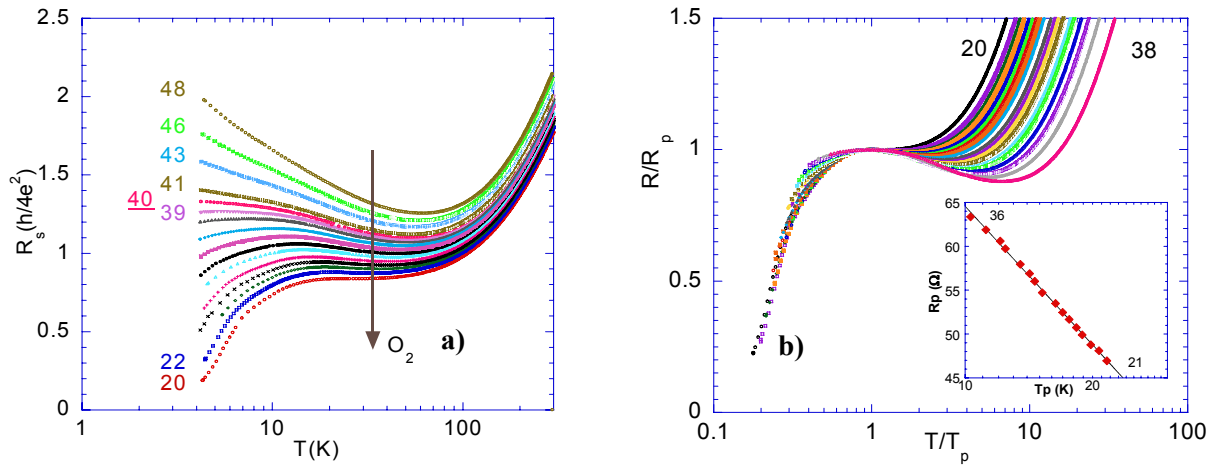

**Figure 1.** (a) Resistance per square *vs logT* with decreasing oxygen contents from states 20 to 48. The state 40 indicates the location of the SIT. (b) Superimposition of the low T part of *R* renormalized by $R_p$ and T by $T_p$. In insert, *logT* variation of $R_p(T_p)$ for the states 21-36.

When going from the initial oxidized superconducting state (10) to the most underdoped state (54), it is found that the Hall number, $n_H$, measured at 300K, decreases linearly as a function of the conductivity at 300K, $\sigma_{300K}$. The parameter $\sigma_{300K}$ is taken as a measure of the doping as in previous work [1, 2]. The linear relationship between $n_H$, and $\sigma_{300K}$ indicates that, through the successive annealing treatments, there is mainly a reduction of the carrier concentration, although some disorder may be introduced. Also the order of magnitude of the hole doping variation between two consecutive states is infinitesimally small. It is found equal to $\Delta n_H = 0.01 \cdot 10^{21}/cm^3$ or $0.0017/Cu$.

In Figure 1(a), which shows the evolution of $R_s(T)$ produced by decreasing the oxygen content through the states 20 to 48, it is seen that at high temperature there is a metallic decrease of the resistance vs T, whatever the doping and even for states where superconductivity is not observed at low temperature. This is in contrast to the behaviour of the resistance of very disordered conventional superconducting films, like amorphous TiN thin films [7], where, close to the SIT, a non metallic resistance is observed for T≤300K. For the states numbered 20 to 38, it is also seen that when decreasing temperature from 300K, the resistance $R_s(T)$ goes through a minimum value $R_{min}$ at $T_{min}$, and increases up to a maximum resistance $R_p$ at $T_p$ , followed by a superconducting transition at $T_c$. This is the so-called reentrant resistive behavior [3]. While at high temperature, above $T_{min}$, all the curves are very close, below $T_{min}$, the curves starts to diverge away from each other, as shown in Figure 1(a), where the curves are displayed with a *logT* scale.

Following the authors of ref.3, we have replotted these data by renormalizing $R$ by $R_p$ and $T$ by $T_p$. The results are presented in Figure 1(b) for the states 20-38 where a maximum $R_p(T_p)$ can be determined. A superimposition of the low T part of these curves is observed. The values of $R_p(T_p)$, are given in insert of Figure 1(b) and appear to be well described by a *logT* variation, while in ref.3 a power law was obtained.

This scaling indicates that there exists a unique function $f(T/T_p)$ to describe this low temperature dependence of the resistance close to the SIT. This region can be decomposed into two parts. For $T<T_p$, $R(T)$ is dominated by superconducting fluctuations, possibly of the Aslamasov-Larkin type as proposed in [3] and observed in a $Bi_2Sr_2CuO_y$ thin film [5] (which are only function of $T/T_c$. whatever the doping). For $T>T_p$, the weakly insulating behavior may be described by a *logT* increase of the resistance. Such a logarithmic increase of $R(T)$ is commonly observed close to the critical doping in cuprates and is better revealed under high magnetic field [8]. However its interpretation is not clear yet. Weak localization and e-e interaction effects (often cited), as observed in 2D conventional superconducting thin films [9], leads to a *logT* behavior of the conductivity with a specific slope not verified here. The Kondo effect will also conduct to a *logT* behavior followed by a saturation at low T not really evidenced yet. Finally it must be noted that the scaling is no longer well satisfied above state 38. One reason is the difficulty to determine $T_p$. But also some influence of disorder may enter into play to further reduce $T_c$.

Considering now the curves $R_s(T)$ in the insulating region, their behavior is similar to the one reported in the case of a film of $Bi_2Sr_{1.6}La_{0.4}CuO_y$ [2]. Namely below $T_{min}$, there is first a region where there is a thermally excited behaviour, below which a 2D variable range hopping (VRH) behaviour takes place. The former is characterized by a variation of the form $exp(E_a/k_bT)$ while the latter is of the form $R_0 exp(T_0/T)^{1/3}$. We obtain similar values for $E_a$ and $T_0$ for $x=0.6$ and $x=0.4$.

This study allowed us to establish a phase diagram displaying the variation with doping, given by $\sigma_{300K}$, of the characteristic temperatures $T_{min}$, $T_p$ and $T_{cross}$, and $T_c$. ($T_{c\ derive.}$ and $T_c(R=0)$), shown in Figure 2. The temperature $T_{cross}$ is the temperature where the derivative $dR/dT$ is minimum, while $T_{min}$ and $T_p$ are the temperatures at which extrema of $R(T)$ take place ($dR/dT = 0$). The temperature $T_{c\ deriv}$ has been obtained from the maximum of the derivative at the transition. Its variation can be adjusted to a parabolic function (solid line). $T_p$ vs 300K follow approximately the same type of behaviour, although a more rapid decrease than parabolic is observed close to the SIT, which could be attributed to an effect of disorder. Both curves appears to converge to zero around $\sigma_{300K} = 0.62 \pm 0.01 (m\Omega*cm)^{-1}$.

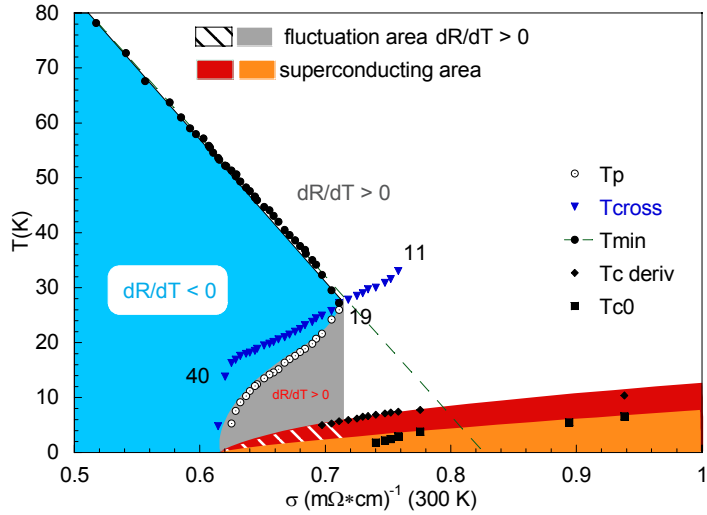

**Figure 2.** The phase diagram displaying the variation with doping, given by $\sigma_{300K}$, of the characteristic temperatures $T_{min}$, $T_p$ and $T_{cross}$, and $T_c$. ($T_{c\ derive.}$ and $T_c(R=0)$)

This result implies that the SIT appears to take place for the states (40-41) (see Fig. 1). Below $T_p$ the system is essentially dominated by superconducting fluctuations. Besides, the temperature $T_{cross}$, which also decreases with decreasing doping in a way similar to that of $T_{c\ deriv}$ and signals the change of curvature of $R(T)$ (from >0 to <0 with decreasing $T$) detects the onset of the influence of superconducting fluctuations on $R(T)$. In contrast the temperature $T_{min}$ increases linearly with decreasing doping. Both $T_{min}$ and $T_p$ appears for $\sigma_{300K} = 0.72$ $(m\Omega*cm)^{-1}$, below which a phase segregation is supposed to take place. In the wedge–shape region delimited by $T_p$ and $T_{min}$ strong Coulomb interaction effect due to strong electronic correlations and/or disorder (for instance in the distribution of oxygen) dominates ($dR/dT<0$). Above $T_{min}$ thermal fluctuations washes out the phase segregation and the metallic-like behavior ($dR/dT>0$) is recovered.

In conclusion, we have observed the reentrant resistive region in a $Bi_2Sr_{1.6}La_{0.6}CuO_y$ thin film by a very fine tuning of the film oxygen content close to the SIT. A scaling of $R(T)$ curves can be observed at low T by renormalizing $R$ and $T$ by $R_p$ and $T_p$ when a peak in $R(T)$ is observed. It may correspond to a "granular" region resulting from electronic phase segregation in hole rich and hole poor region. It is interesting to note that the granular character of $Bi_2Sr_2CuO_y$ has been invoked also to explain the low values (compared to theoretical expectations) of the characteristic electric fields able to destroy the superconducting fluctuations [5]. A phase diagram (T, doping) is established for this compound.

**Acknowledgments**
We wish to acknowledge L. Fruchter and F. Bouquet, Marc Gabay for stimulating discussions.
I. Matei acknowledge the E.C. for a Marie Curie fellowship, contract n°MEST-CT-2004-514307.